\documentclass{INTERSPEECH2023}
\usepackage{xcolor}
\usepackage[hang,flushmargin]{footmisc}
\usepackage{threeparttable}
\interspeechcameraready 

\title{Unified Modeling of Multi-Talker Overlapped Speech Recognition\\and Diarization with a Sidecar Separator}

\name{Lingwei Meng$^1$, Jiawen Kang$^1$, Mingyu Cui$^1$, Haibin Wu$^2$, Xixin Wu$^1$, Helen Meng$^1$}

\address{
  $^1$ Dept. of Systems Engineering \& Engineering Management, The Chinese University of Hong Kong\\
  $^2$ Graduate Institute of Communication Engineering, National Taiwan University}

\email{\{lmeng, jwkang, mycui, wuxx, hmmeng\}@se.cuhk.edu.hk, f07921092@ntu.edu.tw}

\begin{document}

\maketitle
 
\begin{abstract}
Multi-talker overlapped speech poses a significant challenge for speech recognition and diarization. Recent research indicated that these two tasks are inter-dependent and complementary, motivating us to explore a unified modeling method to address them in the context of overlapped speech. 

A recent study proposed a cost-effective method to convert a single-talker automatic speech recognition (ASR) system into a multi-talker one, by inserting a \textit{Sidecar} separator into the frozen well-trained ASR model.
Extending on this, we incorporate a diarization branch into the Sidecar, allowing for unified modeling of both ASR and diarization with a negligible overhead of only 768 parameters. 
The proposed method yields better ASR results compared to the baseline on LibriMix and LibriSpeechMix datasets. 
Moreover, without sophisticated customization on the diarization task, our method achieves acceptable diarization results on the two-speaker subset of CALLHOME with only a few adaptation steps. 

% The proposed approach demonstrates a potential paradigm for cost-effective unified modeling of multi-talker overlapped speech recognition and diarization.
\end{abstract}


\section{Introduction}
% 大纲
% multi-talker overlapped speech 对语音处理领域的很多任务带来了挑战，比如ASR和diarization

Multi-talker (or multi-speaker) overlapped speech has presented significant challenges for many tasks in speech techniques, such as automatic speech recognition (ASR) and diarization \cite{li2022recent}. Although recent years have seen progress in addressing this scenario, these studies tend to be conducted independently, isolated within their respective fields. 

% ASR
In the field of multi-talker overlapped speech recognition, two dominant paradigms have emerged: cascade architectures and fully end-to-end models, both with their drawbacks \cite{li2022recent}.
The cascade architectures, where an ASR module follows a speech separation module, require joint training \cite{settle2018end, li2021real}, and may cause performance degradation in the modules' original domains. 
In contrast, the carefully customized end-to-end models typically necessitate extensive training efforts from scratch \cite{kanda2020serialized, lu2021streaming, kanda2022streaming1,lu2021streaming1}, and fail to capitalize on significant achievements made in common single-talker ASR.
To overcome the limitations of current multi-talker ASR paradigms, a novel \textbf{Sidecar} approach has been recently proposed \cite{meng2023sidecar}. 
It involves loosely coupling with well-trained single-talker ASR models, which are then efficiently adapted for multi-talker scenes, without altering the original model’s parameters. 
Specifically, the Sidecar approach converts a single-talker wav2vec 2.0-based ASR model into a multi-talker ASR model, by plugging a Sidecar separator between two lower encoder layers. This integration utilizes speech separation techniques to effectively disentangle overlapped acoustic embeddings from multiple speakers, thus equipping a standard ASR system to manage multi-talker ASR at a minimal cost.
This research inspires us that the representations hierarchically extracted by ASR encoder \cite{pasad2021layer,shim2021understanding} can be leveraged to handle multiple tasks in a cost-effective manner.

% This work was inspired by two aspects: (1) Studies on layer-wise analysis of ASR encoders found that the lower layers encode more acoustic-related features and the upper layers more linguistic \cite{shim2021understanding, pasad2021layer}; (2) Speech separation task typically only involves low-semantic-level operations \cite{wang2018supervised, luo2019conv}. 
% Enlightened by the above findings, the authors assumed that there exists a lower suitable location between the ASR encoder's two layers for separating the overlapped embedding by drawing on speech separation techniques, and empowering a common ASR system to deal with multi-speaker ASR at a low cost. 
% It picked up a well-trained ASR model and plugged a Sidecar separator into two lower layers of the ASR encoder. In the implementation, it employed a wav2vec 2.0 base-based ASR model \cite{baevski2020wav2vec}, a Conv-TasNet-like Sidecar, with only CTC loss. With very few parameters trainable and limited training efforts, the approach achieved promising performance on multi-talker overlapped speech recognition.

In the diarization field, end-to-end modeling methods have emerged as a promising alternative in recent years \cite{fujita2019eendlstm, fujita2019eendsa, huang2020speaker, Horiguchi2020}, demonstrating superiority in handling overlapped speech compared to conventional pipelines based on speaker embedding clustering \cite{shum2013unsupervised,garcia2017speaker,zheng2022cuhk}.
Although limited in number, some recent studies have investigated joint multi-talker speech recognition and diarization modeling, and indicated that the two tasks are inter-dependent and complementary \cite{park2022review}.
Methods such as \cite{shafey2019joint}, \cite{mao2020speech}, and speaker-attributed ASR \cite{kanda2020joint, kanda22b_interspeech} predict ASR transcriptions alongside sentence-level speaker identification. However, these works do not explicitly output timestamps for speaker activity boundaries.
\cite{kanda2019simultaneous} proposes to iteratively apply an external speaker embedding extractor and a target-speaker ASR model; and similarly the pipeline in \cite{kanda2022transcribe} requires an external pre-trained speaker embedding extractor. They showed promising results in ASR with additional timestamps, but their systems are intricate and not truly unified in modeling the ASR and diarization tasks.
RPN-JOINT \cite{huang2022joint} is a cascade architecture consisting of a diarization module followed by an ASR module with shared lower blocks. The modules are pre-trained in their respective domains and subsequently fine-tuned jointly.
However, since the majority of the two modules remain separate, the overall architecture can still be quite cumbersome.

Encouraged by the capability of the Sidecar in separating embeddings, we aim to extend the prospects of its application for low-cost, end-to-end unified modeling of ASR and diarization. Instead of deploying individual ASR and diarization modules, we hypothesize that a unified backbone could foster knowledge sharing between the two tasks. As shown in Figure \ref{fig:sidecarv2}, building upon the Sidecar approach, we incorporate a diarization branch with merely 768 additional parameters, thereby enabling the unified modeling of both ASR and diarization tasks with negligible computational overhead. The total number of trainable parameters is 8.7 M (8.4\% of all parameters) for the two-speaker model and 8.8 M (8.5\% of all parameters) for the three-speaker model. 
The contributions of the proposed method are threefold:

\begin{itemize}
\item[$\bullet$] We propose a pioneering framework for unified modeling multi-talker ASR and diarization tasks. Exploiting the frozen well-trained ASR model, this approach only contains a small number of trainable parameters hence easy to implement.
% Leveraging the inter-dependency between ASR and diarization tasks, our proposed method outperforms the original Sidecar scheme (the baseline in this work)  on both two-speaker and three-speaker overlapped speech recognition tasks. While the original Sidecar scheme was only implemented on the two-speaker datasets.
\item[$\bullet$]
Leveraging the inter-dependence between ASR and diarization tasks, our proposed method outperforms the original Sidecar scheme for both two- and three-speaker overlapped speech recognition tasks. We demonstrate that this strategy is not only feasible with wav2vec 2.0 backbone, but also with data2vec 2.0 \cite{baevski2022data2vec, baevski2022efficient} - yielding even better results for ASR. 
\item[$\bullet$] Furthermore, without any sophisticated customization on diarization task, our proposed method achieves acceptable performance on the two-speaker subset of CALLHOME with only a few adaptation steps.
\end{itemize}
% \vspace{-0.1cm}
% % diarization
% Diarization, also known as “who spoke when,” is the process of distinguishing and classifying the different speakers present in audio. It is considered a pre-processing step for ASR from a conventional perspective, and also plays an important role in information retrieval from speech signals such as meetings and conversations.

\noindent
We believe that the proposed method holds the potential for a cost-effective solution for the unified modeling of multi-talker overlapped speech recognition and diarization.

\section{Unified Modeling of Multi-Talker Speech Recognition and Diarization with Sidecar}
The proposed approach comprises three main components: a well-trained single-talker ASR model with the parameters frozen, a Sidecar separator with diarization branch, and the training objective. Figure \ref{fig:sidecarv2} illustrates that the Sidecar with diarization branch is inserted between two ASR encoder layers, aided by one convolutional layer on each side, creating a multi-talker ASR and diarization unified modeling system. 
The model is optimized  with permutation invariant training (PIT) \cite{yu2017permutation} for connectionist temporal classification (CTC) loss \cite{graves2006ctc}, and using the same permutation for the diarization loss.

No lexicons or language models are involved in this work.

\subsection{Well-trained single-talker ASR model}
An end-to-end ASR model typically comprises an encoder that converts waveform or acoustic features into high-level representations and a decoder that models these representations into language tokens. However, training such a model from scratch can be time-consuming and challenging, especially in multi-talker environments. As indicated in \cite{meng2023sidecar}, a Sidecar separator can re-purpose existing single-talker models for multi-talker overlapped speech recognition with low cost. 

As a well-known pre-trained speech representation model based on self-supervised learning (SSL), wav2vec 2.0 \cite{baevski2020wav2vec} has gained significant attention in the field of ASR. To adhere to a widely accepted speech representation model, we utilize a well-trained wav2vec 2.0 base-based ASR model, as used in the original Sidecar paper \cite{meng2023sidecar}. Additionally, to validate the feasibility and generality of the Sidecar approach, we also employ a data2vec 2.0 base-based ASR model as another backbone for comparison \cite{baevski2022efficient}.  The two models differ primarily in two ways: (1) they are pre-trained using different protocols, resulting in variations in the encoded representations, and (2) data2vec 2.0 biases the query-key attention scores with a penalty proportional to their distance. 

Both ASR models are well-trained and comprise a CNN feature extractor, a Transformer encoder, and a fully-connected layer as the decoder. Specifically, the model takes waveform as input and extracts acoustic features using a seven-layer CNN feature extractor. The extracted features will be fed into the 12-layer Transformer encoder to generate high-level representations. Following the paradigm outlined in \cite{baevski2020wav2vec} and \cite{baevski2022efficient}, we use only a fully-connected layer as the decoder for letter-level prediction. We directly utilize fairseq's official released model parameters \cite{ott2019fairseq}, and denote them as \emph{W2V-CTC} and \emph{D2V-CTC} throughout subsequent sections.

\subsection{Sidecar separator}
Enlightened by the findings that the ASR encoder captures more acoustic information in its lower layers and more linguistic information in the upper layers \cite{pasad2021layer, shim2021understanding}, a recent study proposes using a Sidecar separator to address multi-talker speech recognition, drawing on methodologies in speech separation \cite{meng2023sidecar}. 

\begin{figure}[t]
  \centering
  \centerline{\includegraphics[height=1\linewidth]{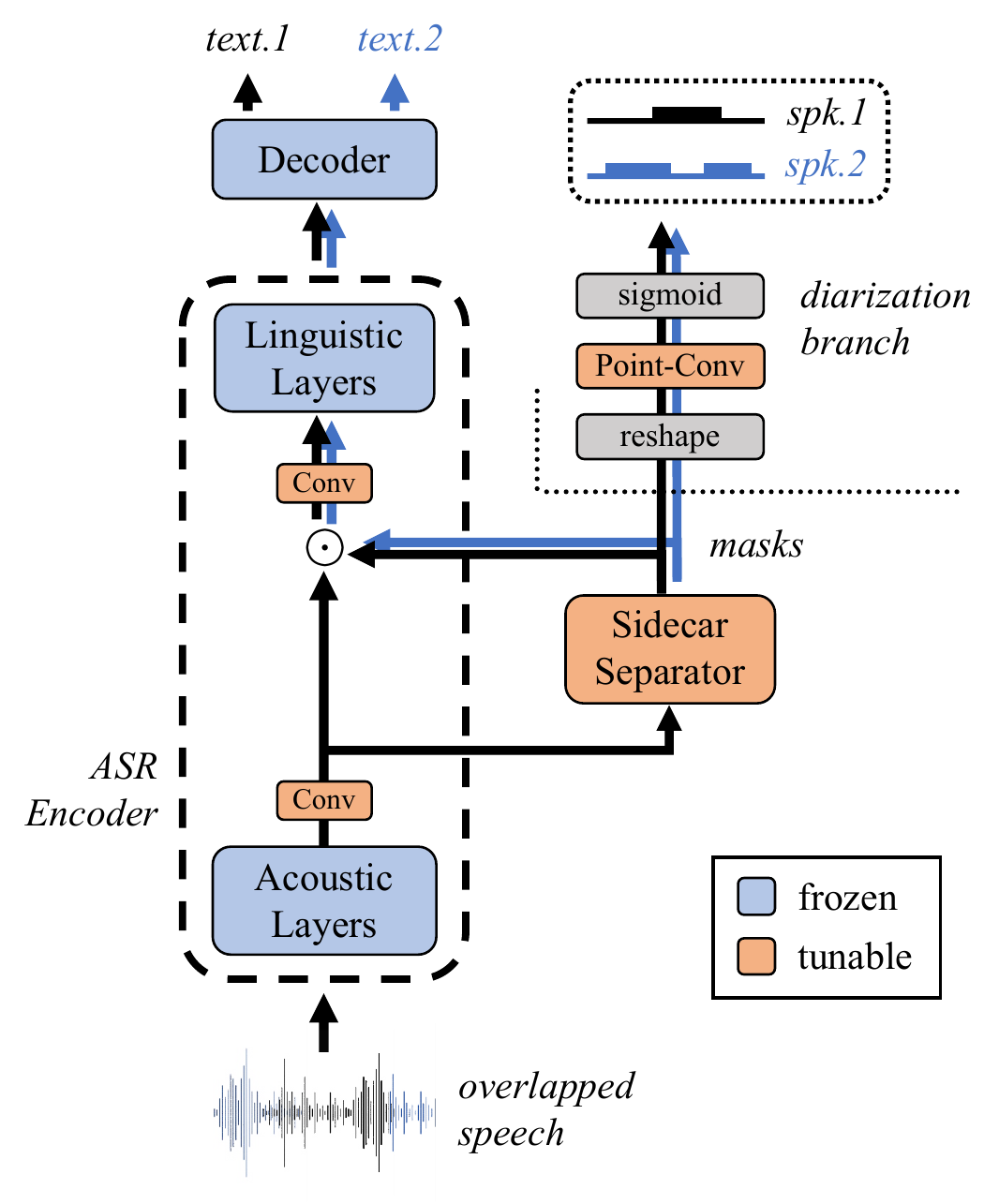}}
  \vspace{-0.3cm}
  \caption{The proposed strategy plugs a Sidecar separator with a diarization branch into a frozen well-trained single-talker ASR model, enabling it unified modeling for multi-talker overlapped speech recognition and diarization. The Sidecar is with Conv-TasNet-like architecture.}
  \label{fig:sidecarv2}
   \vspace{-0.5cm}
\end{figure}

The Sidecar separator is a temporal convolutional network that comprises stacked 1-D dilated convolutional blocks similar to Conv-TasNet \cite{luo2019conv}. This design allows the Sidecar to model long-term dependencies of acoustic embeddings while maintaining a small size. As shown in Figure \ref{fig:sidecarv2}, a 3-kernel-size 1-D convolutional layer is employed on each side of the Sidecar to filter the input-mixed and output-separated embeddings. Following the design proposed in \cite{meng2023sidecar}, we plug the Sidecar between the second and the third encoder layers as a compromise in semantics.

During the forward process, the mixed speech embedding generated by the preceding layer is filtered through a convolutional layer and then fed into the Sidecar to synthesize speaker-dependent masks. These masks are used to element-wise multiply the filtered mixed speech embedding, and the resulting product is further adjusted with another convolutional layer to obtain separated embeddings. These embeddings, corresponding to different speakers, are concatenated onto the batch dimension for parallel processing and transcription into text.

\subsection{The diarization branch and processes for diarization}
\label{subsec:method.diari}
Furthermore, we incorporate a diarization branch into the Sidecar to enable the unified modeling of speech recognition and diarization. As illustrated in Figure \ref{fig:sidecarv2}, the main component of the branch is a point-wise 2-D convolutional layer.

In the forward process, the speaker-dependent masks, as shown in Figure 1, generated by the Sidecar possess a tensor shape of $(B\times S, C, T)$ where $B$ denotes batch size, $S$ denotes the number of speakers, $C$ denotes the number of channels, and $T$ denotes time frames. Each time frame spans a duration of 20 ms. Within the diarization branch, these masks are first reshaped to $(B,S,C,T)$ before being transposed to $(B,C,S,T)$. The reshaped-and-transposed masks then go through a point-wise 2-D convolutional layer, which is the only trainable layer in the branch having $C$ parameters, to generate a tensor with a shape of $(B, 1, S, T)$. After squeezing on the second dimension and applying a sigmoid activation function, frame-wise predictions for each speaker's speech activities $D$ are synthesized with a shape of $(B,S,T)$. For every speaker $s$ in $S$ and for each time frame $t$ in $T$, if the element value is greater than 0.5, we consider that speaker $s$ was activated on the time frame $t$. This will yield the results for diarization.

\begin{figure}[t]
  \centering
  \centerline
  {\includegraphics[height=0.32\linewidth]{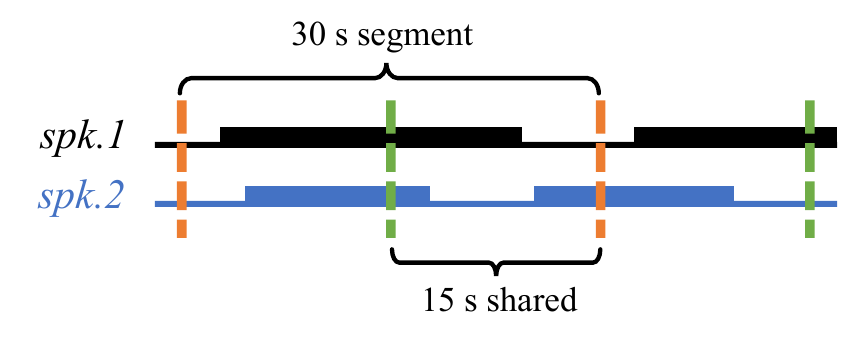}}
  \vspace{-0.3cm}
  \caption{The pre-process for diarization on CALLHOME. The interval between the green and orange dashed lines is the shared interval for the two segments.}
  \label{fig:dairi}
  \vspace{-0.5cm}
\end{figure}

During the training phase on LibriMix and LibriSpeechMix datasets, the model is fed with complete utterances. However, during the adaptation and inference for diarization on CALLHOME (Section \ref{subsec:exp.diari}), we segment the utterances to ensure the alignment with real-world diarization scenarios and guarantee its practicality. As depicted in Figure \ref{fig:dairi}, we divide each utterance into several 30-second segments that share a common 15-second interval between every two adjacent segments. For shared parts, we calculate Euclidean distance for different speaker permutations between segment tensors $D$ and select one with minimum distance to modify the speaker arrangement in subsequent segments. Afterward, we average the element values of adjacent segments' shared parts.
Note that this segmenting-and-permuting process is not utilized in the experiments mentioned in Section \ref{subsec:asr}.

% \subsection{Processes for diarization inference}
\vspace{-0.2cm}

\subsection{Training and adaptation objectives}
During training on LibriMix and LibriSpeechMix datasets, both CTC loss and diarization loss require a permutation for the speaker order to be assigned  to address the label ambiguity issue \cite{yu2017permutation}. To explicitly construct the inter-dependence between the two tasks, the permutation is determined by permutation invariant training (PIT) based on CTC loss, and then is assigned for diarization loss.
% During training on LibriMix and LibriSpeechMix datasets, permutation invariant training (PIT) is used to determine the speaker permutation based on CTC loss. 
% To construct the inter-dependence between the two tasks. Subsequently, the permutation that yields minimum CTC loss is utilized for computing diarization loss as well. 
% After the speaker permutation is determined, 
The diarization loss is to calculate the mean squared error (MSE) between the predicted speaker activities $D$ and the diarization ground truth.
 At last, the final objective function is the sum of PIT-CTC loss and corresponding diarization loss multiplied by a coefficient $\lambda$.

However, when we adapt the model for CALLHOME, we solely employ diarization loss and determine speaker permutation relying on the strategy outlined in Section \ref{subsec:method.diari}.

\vspace{-0.1cm}

\section{Experimental Setup}
\vspace{-0.1cm}
\subsection{Datasets}
The experiments are performed on two benchmark datasets for multi-talker ASR (LibriMix \cite{cosentino2020librimix} and LibriSpeechMix \cite{kanda2020serialized}), and the two-speaker subset of a real-world dataset for diarization (CALLHOME).  Although LibriMix and LibriSpeechMix datasets were not specifically designed for the diarization task, we also present diarization results on them.
% Both LibriMix and LibriSpeechMix datasets are simulated from LibriSpeech.

\noindent
\textbf{LibriSpeechMix}. 
The utterances are simulated with the mixtures of two or three speakers from LibriSpeech.
Only standard official dev and test sets are published.
Our training set is homemade from the 960-hour LibriSpeech training dataset (LS-960) referring to the protocol established in \cite{kanda2020serialized}.
LibriSpeechMix randomly samples a delay time for the second and the third utterances, so the mixture is partially overlapping.

\noindent
\textbf{LibriMix}.
The dataset simulates audio mixtures using a combination of two or three speakers sourced from the LibriSpeech-clean corpus.
We focus on its two-speaker-clean subset \emph{Libri2Mix-clean} and three-speaker-clean subset \emph{Libri3Mix-clean}. 
The mixtures are made in a left-aligned style. Thus, the shorter source speech will be entirely overlapped by the longer one from the start, which challenges the model more than LibriSpeechMix in separating overlaps.

\noindent
\textbf{CALLHOME}. We evaluate the proposed method on the diarization task with CALLHOME, which is a benchmark dataset consisting of spontaneous multilingual telephone conversations, as one part of the 2000 NIST Speaker Recognition Evaluation (LDC2001S97). We take its two-speaker subset and split it into an adaptation set of 155 recordings and a test set of 148 recordings, following the same partition protocol with EEND \cite{fujita2019eendlstm, fujita2019eendsa} and Kaldi\footnote{https://github.com/kaldi-asr/kaldi/tree/master/egs/callhome\_diarization/v2} \cite{povey2011kaldi}. The average duration is 73.1 seconds.

\subsection{Model settings}

\noindent
\textbf{Well-trained single-talker ASR model}. 
We utilize the well-trained W2V-CTC and D2V-CTC as the backbone, respectively. To ensure consistency with \cite{baevski2020wav2vec} and \cite{baevski2022efficient}, we directly employ the official released model weights by fairseq\footnote{https://github.com/facebookresearch/fairseq}  \cite{ott2019fairseq}. Both models are pre-trained on unlabeled LS-960 data and subsequently fine-tuned on labeled LS-960 using CTC loss. The resulting well-trained models are then frozen for use in our experiments.

\noindent
\textbf{Sidecar separator}.
Drawing inspiration from \cite{luo2019conv}, the Sidecar separator implements a sequence of $K$ temporal convolutional blocks with dilation rates spanning from 1 to 2$^{K-1}$, repeated up to $K$ times. To align with the protocol in \cite{meng2023sidecar}, we set $K=8$ and $R=3$, remove skip-connection paths within the convolutional blocks, and substitute the final sigmoid activation with ReLU. The Sidecar has 128 bottleneck channels and 768 input/output channels. The Sidecar is plugged between the second and the third transformer layers, which mirrors that of \cite{meng2023sidecar}.

\noindent
\textbf{Diarization branch}.
As shown in Figure \ref{fig:sidecarv2}, the speaker-dependent masks predicted by Sidecar will be fed into the diarization branch. 
The main component of the branch is a point-wise 2-D convolutional layer with an input-channel size of 756 and output-channel size of 1, featuring a kernel size of $(1, 1)$ with stride $1$. As such, this branch only requires an additional 768 parameters. As is typical, we use a collar of 250 ms when evaluating the DER performance \cite{fujita2019eendlstm, fujita2019eendsa}.

\noindent
\textbf{Training settings}. 
With W2V-CTC or D2V-CTC frozen, the models only have 8.7 M trainable parameters (8.4\% of all parameters) for the two-speaker experiments and 8.8 M trainable parameters (8.5\% of all parameters) for the three-speaker experiments. We set the coefficient of diarization loss $\lambda$ to 0.01.
Adhering to the settings in \cite{meng2023sidecar}, we optimize the proposed models using a 2e-4 learning rate with a three-stage schedule and Adam optimizer, for at most 100 k updates.
It takes about 7 hours for two-speaker models and 9 hours for three-speaker models with 8 NVIDIA V100 GPUs, thanks to Sidecar’s small size and the ejection start provided by the well-trained ASR model.

In the following, we denote the proposed models as \textit{W2V-Sidecar}, \textit{W2V-Sidecar-DB}, \textit{D2V-Sidecar}, \textit{D2V-Sidecar-DB}, where “\textit{-DB}” denotes “with diarization branch”. Note that \textit{W2V-Sidecar} is identical to the implementation in \cite{meng2023sidecar}, serving as the baseline in this work. Permutations with minimum errors are used to compute word error rates (WERs).

\section{Results and Discussion}

\subsection{Results on LibriMix and LibriSpeechMix datasets}
\label{subsec:asr}
\vspace{-0.1cm}
The models (b) and (d) in Table \ref{tab:asr} are optimized with CTC and diarization loss, enabling the unified modeling of multi-talker ASR and diarization. 

\noindent
\textbf{ASR.} The ASR results of the four models on two- and three-speaker LibriMix and LibriSpeechMix datasets are presented in Table \ref{tab:asr}. Aligning with our previous hypothesis, the enhanced model (b) with a diarization branch consistently outperforms the baseline (a) \cite{meng2023sidecar} across all four datasets, benefiting from the inter-dependent and complementarity of the two tasks.
Additionally, models utilizing D2V as a backbone generally outperform those with W2V. We attribute the boost in performance of models (c) and (d) to the better representations learned in its pre-training phase as illustrated in \cite{baevski2022efficient}. The diarization branch's performance enhancement on the LibriSpeechMix-2pk is relatively modest, and we contend that this is due to the dataset being relatively simple for ASR owing to its lower two-speaker overlap rate. This study achieves the state-of-the-art performances on Libri2Mix, and it is the first to report ASR results on Libri3Mix in the field. However, WERs on Libri3Mix remains high due to shorter source speeches being completely overlapped by longer ones from the outset, rendering it more challenging for multi-talker ASR.

\noindent
\textbf{Diarization.} Although LibriMix and LibriSpeechMix datasets were not created for diarization purposes, we include the diarization results on them in Table \ref{tab:libridiari} to offer a more comprehensive perspective. 
As a similar trend to Table \ref{tab:asr}, D2V backbone achieves better performance on the diarization task compared to W2V.
Note that while left-aligned-style generated LibriMix dataset poses greater challenges for ASR than LibriSpeechMix, it proves easier for the diarization task since it only requires predicting speaker activity timestamps on the right side.

% \subsection{Diarization results on LibriMix and LibriSpeechMix}

\vspace{-0.1cm}
\subsection{Diarization results on CALLHOME dataset}
\label{subsec:exp.diari}
\vspace{-0.1cm}
% 突出对比EEND没有复杂定制
%
To demonstrate its practicality, we conducted the evaluation of the proposed method on the real-world CALLHOME dataset under realistic settings. 
Specifically, we adapt W2V-Sidecar-DB and D2V-Sidecar-DB, which have been trained on Libri2Mix, to the two-speaker subset of CALLHOME. During adaptation and inference, the segmenting-and-permuting process is employed to align with real-world diarization scenes, as discussed in Section \ref{subsec:method.diari}. As shown in Table \ref{tab:diari}, in comparison to EEND models that are carefully designed for diarization and trained on datasets tailored for this purpose, our method delivers satisfactory performance with just 8.7 M trainable parameters and a few adaptation steps, demonstrating the effectiveness and flexibility of our approach with limited resources.

We observed that the proposed method can achieve better or comparable performance on MI and FA, which are metrics related to voice activity detection, but falls behind in CF. We argue that this is because the used Libri2Mix dataset is clean and designed for ASR purpose with only 1,172 speakers, far fewer than the 5,743 speakers of the carefully crafted dataset used by EEND.
We anticipate significant improvement by training it with data simulated specifically for diarization purposes as done by \cite{fujita2019eendlstm,fujita2019eendsa}.

% To showcase its practicality, we conducted an evaluation of the proposed method on the real-world CALLHOME dataset under realistic settings. Specifically, we adapted W2V-Sidecar-DB and D2V-Sidecar-DB, which were trained on Libri2Mix, to the two-speaker subset of CALLHOME. During adaptation and inference, we employed the segmenting-and-permuting process to align with real-world diarization scenes as discussed in Section \ref{subsec:method.diari}. As demonstrated in Table \ref{tab:diari}, our approach delivers satisfactory performance with just 8.7 M trainable parameters and a few adaptation steps compared to EEND models that are carefully designed for diarization and trained on datasets tailored for this purpose. This highlights the effectiveness and flexibility of our approach even when resources are limited. We observed that our proposed method can achieve better or comparable performance on MI and FA metrics related to voice activity detection but falls behind in CF due to using a cleaner dataset designed for ASR purposes with only 1,172 speakers compared to EEND's use of 5,743 speakers. However, we argue that training with data simulated for diarization purposes as done by \cite{fujita2019eendlstm,fujita2019eendsa} or employing a more powerful Sidecar structure could significantly improve results in this area.
\vspace{-0.1cm}
\subsection{Limitations and future work}
\vspace{-0.1cm}
While this work is innovative in unifying the modeling of speech recognition and diarization, several limitations remain. Firstly, due to considerations of simplicity and comparability with previous work \cite{meng2023sidecar}, the model's performance on diarization is acceptable but restricted by its training strategy. 
However, access to more suitable datasets and training schemes may yield improved performance. Secondly, the system requires pre-defining the maximum number of speakers addressed, which could potentially be resolved by maintaining a speaker embedding bank in the future. Lastly, the current model does not tackle the "who spoke when and what" issue. Nevertheless, we believe this work has pointed the direction toward a potential solution to it. 

\vspace{-0.2cm}
\section{Conclusion}
\vspace{-0.05cm}
A recent study proposed a low-cost approach for converting a single-talker ASR system to a multi-talker one, by plugging a Sidecar separator into a fixed well-trained common ASR model. Extending on this approach, we incorporate a diarization branch into the Sidecar with only 768 additional parameters, allowing for unified modeling of both multi-talker overlapped speech recognition and diarization tasks with a low training cost. 

With very few parameters (8.7 M for the two-speaker model, and 8.8 M for the three-speaker model) requiring tuning, the proposed approach outperforms the original Sidecar scheme on ASR tasks for LibriMix and LibriSpeechMix datasets. Furthermore, without sophisticated customization on the diarization task, the proposed method achieves potential diarization results on the two-speaker subset of the real-world CALLHOME dataset, with only a few adaptation steps.

\begin{table}[!t]
  % \footnotesize
% \vspace{0.1cm}
  \caption{ASR performance on the test sets of LibriMix and LibriSpeechMix. Evaluated by WER (\%). “-DB” refers to “with diarization branch”}
\vspace{-0.25cm}
  \label{tab:asr}
  \centering 
    \begin{threeparttable}
  {
  \begin{tabular}{lcccc}

    \toprule
    & \multicolumn{2}{c}{\textbf{LibriMix}} & \multicolumn{2}{c}{\textbf{LibriSpeechMix}} \\
    \cmidrule(r){2-3} \cmidrule(r){4-5}
    \textbf{System} & 2spk & 3spk   &    2spk & 3spk     \\
    \midrule
   (a) W2V-Sidecar\textsuperscript{\dag} \cite{meng2023sidecar} & 10.36  &  35.22   & 7.56  & 13.87    \\
   (b) W2V-Sidecar-DB     & 9.88 & 34.38 & 7.53 &  12.93  \\
  \midrule
  (c)   D2V-Sidecar                 & 10.11   & 34.84  & 7.61  &  12.56 \\
   (d)   D2V-Sidecar-DB   & \textbf{9.69} & \textbf{33.91}  & \textbf{7.49} & \textbf{11.94} \\
    \bottomrule
    \end{tabular}
    }

\begin{tablenotes}

\item\textsuperscript{\dag} Serving as the baseline in this work.
\end{tablenotes}  
\end{threeparttable}
\vspace{-0.25cm}
\end{table}

\begin{table}[!t]
  % \footnotesize
% \vspace{0.1cm}
  \caption{Diarization performance on the test sets of LibriMix and LibriSpeechMix. Evaluated by DER (\%).}
\vspace{-0.25cm}
  \label{tab:libridiari}
  \centering 
  \setlength{\tabcolsep}{2.9mm}
  {
  \begin{tabular}{lcccc}

    \toprule
    & \multicolumn{2}{c}{\textbf{LibriMix}} & \multicolumn{2}{c}{\textbf{LibriSpeechMix}} \\
    \cmidrule(r){2-3} \cmidrule(r){4-5}
    \textbf{System} & 2spk & 3spk   &    2spk & 3spk     \\
    \midrule
    W2V-Sidecar-DB    & 0.97  &  2.35  & 2.20 & 3.65    \\

    D2V-Sidecar-DB    & 0.91  & 2.14 & 2.12  & 3.47 \\

    \bottomrule
    \end{tabular}
    }
\vspace{-0.2cm}
\end{table}

\begin{table}[!t]
%   \footnotesize
% \vspace{0.1cm}
  \caption{Detailed diarization results on CALLHOME, evaluated by DER (\%), which is the sum of misses (MI), false alarms (FA), and speaker confusions (CF) errors.}
    \vspace{-0.25cm}
  \label{tab:diari}
  \centering 
  \setlength{\tabcolsep}{2.5mm}
  {
  \begin{tabular}{lccc|cc}

    \toprule

    \textbf{System} & \textbf{MI} & \textbf{FA} & \textbf{CF} & \textbf{DER}     \\
    \midrule
    BLSTM-EEND \cite{fujita2019eendlstm}       & - & - &    -   &   23.07   \\
    SA-EEND  \cite{fujita2019eendsa}        & 6.68  & 2.40 & 1.68 &   \textbf{10.66}   \\
    W2V-Sidecar-DB (ours)       & 6.26 &  3.56 &  5.07  & 14.89      \\
        D2V-Sidecar-DB (ours)      & 6.18 & 3.80  & 4.64   & 14.62     \\

    \bottomrule
    \end{tabular}
    }
    \vspace{-0.5cm}
\end{table}

\vspace{-0.15cm}
\section{Acknowledgements}
\vspace{-0.05cm}
This research is partially supported by the HKSARG Research Grants Council’s Theme-based Research Grant Scheme (Project No. T45-407/19N) and by the CUHK Stanley Ho Big Data Decision Analytics Research Centre.

\vfill
\pagebreak

\bibliographystyle{IEEEtran}
\bibliography{mybib}

\end{document}